\documentclass[12pt,a4paper]{article}

\usepackage{amsmath,amssymb,amsfonts}
\usepackage{graphicx}
\usepackage{booktabs}
\usepackage{caption}
\usepackage{geometry}
\usepackage{abstract}
\usepackage{authblk}
\usepackage{tikz}
\usetikzlibrary{quantikz2}
\usepackage{quantikz}
\usepackage{amsmath, amssymb, physics}
\usepackage{physics}
\usepackage{float}

\usepackage{mathrsfs}
\usepackage{makecell}

\usepackage{hyperref}
\hypersetup{
	colorlinks=true,
	linkcolor=blue,
	citecolor=red,
	urlcolor=blue,
}
\title{\textbf{Quantum Circuit Realization of the PPT and CCNR Criteria}\footnote{This research is supported by Natural Science Foundation of China (No. 11761073)}}
\author{
	\begin{minipage}{\textwidth}
		\centering
		Ji-Hui Mo\textsuperscript{},
		Yuan-Hong Tao\textsuperscript{$\dagger$} \quad

		\small
		\textsuperscript{} School of Science, Zhejiang University of Science and Technology, Hangzhou 310023, China 
		
		$\dagger$ Corresponding author: taoyuanhong12@126.com
		
	\end{minipage}
	}
\date{} 

\begin{document}
	\maketitle

	\begin{quote}
		
	    \hspace{1.5em} \textbf{Abstract:}
		The efficient detection of quantum entanglement is a central problem in quantum information processing. This paper systematically proposes a quantum circuit implementation scheme based on the Positive Partial Transpose (PPT) and the Computable Cross-Norm Realignment (CCNR) criteria, providing a complete quantum algorithmic pathway for efficient and computable entanglement detection. By encoding quantum states into specific forms and utilizing SWAP operations, complex matrix operations such as partial transpose and realignment are transformed into executable quantum circuits. Furthermore, by integrating an improved variational quantum singular value decomposition subroutine, the scheme enables the efficient estimation of the trace norm, thereby determining the existence of entanglement. Designed to operate within a hybrid quantum-classical framework, this scheme exhibits excellent scalability and practicality, offering a theoretical tool and methodological support for analyzing the entanglement structure of complex quantum systems on future intermediate-scale quantum devices.

		\hspace{1.5em} \textbf{Keywords:}
		quantum entanglement, PPT criterion, CCNR criterion, quantum singular value decomposition, trace norm, quantum circuit
	\end{quote}

	\section{Introduction}
	Quantum entanglement, as one of the most remarkable phenomena in quantum mechanics, serves not only as a fundamental resource for quantum information science but also as a key signature distinguishing the quantum world from the classical realm. It plays an indispensable role in cutting-edge fields such as quantum computation,quantum communication, and quantum precision measurement\cite{yyu2021,nxzou2021,gisin2007}. Among them,he effective detection of quantum entanglement remains a task of fundamental importance and significant challenge. Its implications are profound, primarily manifesting at three levels, at the foundational theory level:entanglement is key to testing the nonlocality of quantum mechanics, and its verification is a central part of understanding the fundamental principles of the quantum world. at the resource certification level:entanglement is an indispensable resource in quantum communication (e.g., quantum teleportation, quantum key distribution) and quantum computation (e.g., measurement-based quantum computing). The success of these operations directly depends on whether the states used are genuinely entangled; therefore, their reliable certification is a prerequisite for any application. at the system characterization level:as quantum processors scale up, performing complete quantum state tomography on the complex states they generate has become infeasible. Developing entanglement detection tools that do not require full tomography is the essential path forward for characterizing and verifying the properties of large-scale quantum systems.

	In recent years, the rapid development of quantum algorithms has provided new ideas and methods for addressing quantum entanglement criteria. By leveraging the superposition and entanglement properties of qubits, quantum algorithms can achieve exponential speedups over classical algorithms for certain specific problems. Through the design of appropriate quantum algorithms, efficient matrix operations (including matrix multiplication, matrix inversion, determinant calculation, trace computation, etc. \cite{Zenchuk2025,zenchuk2025quantum,zenchuk2024arbitrary,liu2024quantum,liu2025quantum,Zenchuk202424,zenchuk2024matrix,qiwentao2024matrix}) can be performed on quantum states, enabling efficient measurement and analysis, thereby constructing more accurate and powerful quantum entanglement criteria.
	
	Given that quantum computing can surpass classical computing in solving complex problems, a natural question arises: designing quantum algorithms for entanglement detection to avoid quantum state tomography. Currently, there are relatively a few achievements in quantum algorithms for entanglement detection, which can be broadly categorized into two types. The first category comprises NISQ algorithms. Preskill \cite{JP2018} noted that quantum computing is entering the NISQ era, making algorithm development for such devices crucial. While many entanglement detection methods exist, few are tailored for NISQ devices, posing a key challenge. Recent works address this: Refs. \cite{KAMB2006,YZPZ2020,AE2020,AKSI2022,FSSM2022} measure randomly sampled bases, using statistical correlations as entanglement signatures. X. Wang \cite{KWZS2022} combined variational quantum algorithms with quasi-probability decomposition for hybrid entanglement detection. R. Chen \cite{RCBZ2023} implemented Schmidt decomposition on NISQ devices to detect entanglement in mixed states. M. Consiglio et al. \cite{MCTJ2022} introduced a variational separability verifier to find the closest separable state.
	
	The second category of quantum algorithms is based on the controlled-SWAP (c-SWAP) test. In 2020, Foulds et al. \cite{SFVK2021} first utilized the c-SWAP test to develop a quantum algorithm for detecting entanglement in multipartite pure states. Subsequently, references \cite{OPSF2022,RQYD2024,AGYS2019,XMTL2022} applied the c-SWAP test to detect high-dimensional bipartite states, multipartite qubit states, and some special types of entangled optical states. Reference \cite{JLNG2021} proposed a series of methods for efficiently estimating multipartite pure state metrics through parallelized c-SWAP tests.
	
	However, algorithms for entanglement detection based on entanglement criteria remain underdeveloped, as traditional quantum entanglement criteria have certain limitations in practical applications. Moreover, criteria such as the partial transpose criterion and entanglement witnesses often rely on quantum state tomography, whose resource consumption grows exponentially with system size, making them unsuitable for large-scale quantum systems. With the advancement of quantum technology, there is an urgent need to develop novel, efficient, and universal quantum circuits for entanglement criteria.
	
	Among many theoretical tools for entanglement detection, the PPT criterion and the CCNR criterion hold central positions due to their computational feasibility. The PPT criterion, proposed by Peres~\cite{Peres1996} in 1996, states that the partial transpose of any separable state must be positive semidefinite. Subsequently, the Horodecki family~\cite{Horodecki1996} rigorously proved the completeness of this criterion for $2 \otimes 2$ and $2 \otimes 3$ systems. Almost simultaneously, two research groups, Rudolph~\cite{Rudolph2003} and Chen \& Wu~\cite{Chen2002}, independently proposed the realignment criterion, which serves as a "sister" criterion to PPT. Both criteria ultimately reduce to computing the trace norm (the sum of all singular values) of the transformed matrix.

	Currently, there is no complete compilation of PPT and CCNR criteria into a feasible and efficient quantum circuit for entanglement detection. The realization still faces two key requirements: First, dedicated quantum circuits must be designed to simulate the "partial transpose" and "realignment" operations, which are non-local linear operations on high-dimensional density matrices. Second, an algorithm capable of directly estimating the matrix's nuclear norm is needed to align with the ultimate objective of the entanglement criteria.
	
	This paper addresses this gap by proposing a comprehensive quantum circuit scheme for implementing the PPT and CCNR criteria. The main steps of our work include:
	
	\textbf{Operational Quantization:} By partitioning the density operator into blocks and encoding it into specific quantum states, we ingeniously transform the partial transpose and realignment operations into equivalent quantum operations executable on a circuit through the strategic use of SWAP gates.
	
	\textbf{Nuclear Norm Estimation:} We introduce a crucial modification to the existing Variational Quantum Singular Value Decomposition subroutine. By setting uniform weights, we repurpose it from resolving individual singular values to directly and efficiently estimating the nuclear norm of the transformed matrix.
	
	\textbf{Complete Workflow Integration:} The aforementioned components are seamlessly integrated into a quantum-classical hybrid algorithm. This integrated workflow ultimately outputs the nuclear norm value, providing a clear quantitative basis for determining quantum state entanglement.

	The structure of this paper is as follows: Section 2.1 reviews the theoretical foundations of the PPT and realignment criteria; Section 2.2 introduces the variational quantum singular value decomposition (VQSVD) circuit; Section 3.1 details the VQSVD-optimized subroutine for trace norm estimation; Sections 3.2 and 3.3 elaborate on the quantum circuit implementation details for the PPT criterion and the realignment criterion, respectively; Section 3.4 validates the effectiveness of the scheme through a concrete example using Bell states; finally, Section 4 concludes the paper and discusses prospects for future applications.
	
	\section{Preliminaries}
	\subsection{The PPT and CCNR Criteria}  
	The density matrix for a pure state $|\psi \rangle_{AB}$ is defined with respect to the computational basis as:
	\begin{equation}
		\rho_{AB} =|\psi \rangle_{AB} \langle \psi|=\sum_{ijkl} \rho_{ijkl} | i \rangle \langle j | \otimes | k \rangle \langle l |,
		\label{eq:2.2}
	\end{equation}
	(where $| i \rangle \langle j |$ are the indices for subsystem A, and $| k \rangle \langle l |$ are the indices for subsystem B.)
	
	\noindent The partial transpose $\rho_{AB}^{T_B}$ and the realigned matrix $R(\rho_{AB})$ can be respectively expressed as:
	
	\begin{equation}
		\rho_{AB}^{T_B} = \sum_{ijkl} \rho_{ijkl} | i \rangle \langle j | \otimes | l \rangle \langle k |,
		\label{eq:2.3}
	\end{equation}
	
	\begin{equation}
		R(\rho_{AB}) = \sum_{ijkl} \rho_{ijkl} |i\rangle \langle k| \otimes |j\rangle \langle l|
	\end{equation}

	We illustrate the matrix representations of the aforementioned two operations through a concrete example. Consider a two-qubit bipartite state $\rho_{AB}$, where subsystems A and B are both 1-qubit systems with the Hilbert space $\mathcal{H}_A \otimes \mathcal{H}_B$. In the computational basis ${ \ket{00}, \ket{01}, \ket{10}, \ket{11} }$, we express it in a $2 \times 2$ block-matrix form (where each block is a $2 \times 2$ matrix). Then, a general density matrix $\rho_{AB}$ can be written as:

	\begin{equation}
		\rho_{AB} =
		\left(
		\begin{array}{cc|cc}
			\rho_{00,00} & \rho_{00,01} & \rho_{00,10} & \rho_{00,11} \\
			\rho_{01,00} & \rho_{01,01} & \rho_{01,10} & \rho_{01,11} \\
			\hline
			\rho_{10,00} & \rho_{10,01} & \rho_{10,10} & \rho_{10,11} \\
			\rho_{11,00} & \rho_{11,01} & \rho_{11,10} & \rho_{11,11}
		\end{array}
		\right)
		= \begin{pmatrix}
			E & F \\
			G & H
		\end{pmatrix},
		\label{eq:original_matrix}
	\end{equation}
	
	\noindent Performing the partial transpose operation on subsystem $B$ (denoted as $T_B$):
	\begin{equation}
		\rho_{AB}^{T_B} = \begin{pmatrix}
			E^T & F^T \\
			G^T & H^T
		\end{pmatrix}=\left(
		\begin{array}{cc|cc}
			\rho_{00,00} & \rho_{01,00} & \rho_{00,10} & \rho_{01,10} \\
			\rho_{00,01} & \rho_{01,01} & \rho_{00,11} & \rho_{01,11} \\
			\hline
			\rho_{10,00} & \rho_{11,00} & \rho_{10,10} & \rho_{11,10} \\
			\rho_{10,01} & \rho_{11,01} & \rho_{10,11} & \rho_{11,11}
		\end{array}
		\right)
	\end{equation}

	\noindent Applying the realignment operation $\mathcal{R}$ to $\rho_{AB}$ yields a new matrix $\mathcal{R}(\rho_{AB})$ whose elements are given by:
	\begin{equation}
	[\mathcal{R}(\rho_{AB})]_{ik,jl} = \rho_{ij,kl},
	\end{equation}
	where the composite index $(ik)$ labels the rows and $(jl)$ labels the columns. Applying this rule, we obtain the realigned matrix:
	
	\begin{equation}
		\mathcal{R}(\rho_{AB}) =\begin{pmatrix}
			\mathrm{vec}(E)^T \\
			\mathrm{vec}(F)^T \\
			\mathrm{vec}(G)^T \\
			\mathrm{vec}(H)^T
		\end{pmatrix}=
		\begin{pmatrix}
			\rho_{00,00} & \rho_{00,01} & \rho_{01,00} & \rho_{01,01} \\
			\rho_{00,10} & \rho_{00,11} & \rho_{01,10} & \rho_{01,11} \\
			\rho_{10,00} & \rho_{10,01} & \rho_{11,00} & \rho_{11,01} \\
			\rho_{10,10} & \rho_{10,11} & \rho_{11,10} & \rho_{11,11}
		\end{pmatrix}.
		\label{eq:realigned_matrix}
	\end{equation}
	
	\noindent where $\mathrm{vec}(X)$ denotes the column vector formed by stacking the columns of matrix $X$.

	PPT criterion and CCNR criterion are both theoretical extensions of the generalized index–swap theorem. The PPT criterion was states that if a bipartite state $\rho_{AB}$ is separable, then the new matrix which obtained by applying the partial transpose to subsystem $B$$\rho_{AB}^{T_B}$, must be positive semi-definite. The Horodecki family \cite{Horodecki1996} proved that the PPT criterion serves as a necessary and sufficient condition for entanglement detection in $2 \otimes 2$ and $2 \otimes 3$ systems, while for higher-dimensional systems it remains only a necessary condition for separability (i.e., a violation of the criterion is a sufficient condition for entanglement).
	
	C.J. Zhang \cite{Zhang2010} further demonstrated that the PPT criterion can be equivalently expressed as follows: if a bipartite state $\rho_{AB}$ is separable, the matrix $\rho_{AB}^{T_B}$ must satisfy the inequality:
	\begin{equation}
		\|\rho_{AB}^{T_B}\| \leq 1
	\label{eq:2.6}
    \end{equation}
    where $\| \cdot \|$ denotes the trace norm, defined as $\| A \| = \mathrm{Tr} \sqrt{AA^\dagger}$, which equals the sum of the singular values of $A$.

    The CCNR criterion \cite{Rudolph2003,Chen2002} states that if a bipartite state $\rho_{AB}$ is separable, then its realigned matrix $R(\rho_{AB})$ must satisfy the inequality:
    	\begin{equation}
    		\|R(\rho_{AB})\| \leq 1
    		\label{eq:2.6}
    	\end{equation}
    The CCNR criterion is often regarded as a complement to the PPT criterion, as it can detect certain bound entangled states that fail to be identified by the partial transpose criterion.
	\subsection{Variational Quantum Singular Value Decomposition}

	In 2025, A.I. Zenchuk, W.T. Qi, and J.D. Wu \cite{Zenchuk2025} proposed the Variational Quantum Singular Value Decomposition (VQSVD) algorithm, developing a variational quantum computing procedure for singular value decomposition. This procedure directly encodes the target matrix $A$ into a quantum state $|A\rangle$ and utilizes a parameterized quantum circuit to construct and optimize a specific objective function: \( L(\alpha, \beta) = \sum_j q_j \, \text{Re}([U^T(\alpha) A U(\beta)]_{jj}) \) to compute the singular values and singular vectors of the matrix A. This VQSVD subroutine, combined with a classical optimizer, forms a quantum-classical hybrid algorithm. The core workflow and principles of its quantum component are as follows:
	\begin{figure}[H]
		\centering
		\includegraphics[width=0.8\textwidth]{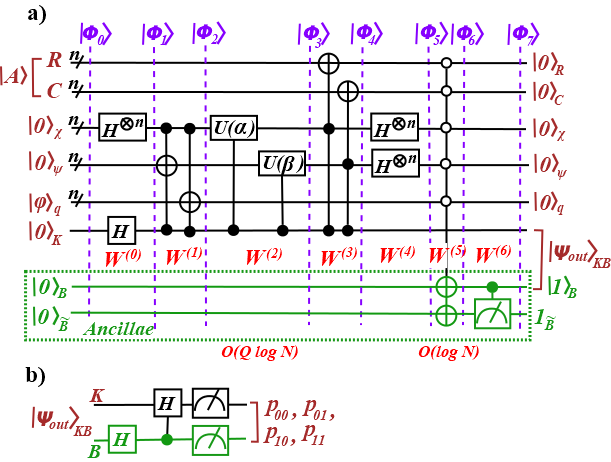}
	\end{figure}
	
	Step 1: \textbf{Matrix Encoding and Initialization}.
	The $N \times N$ matrix $A$ (with $N = 2^n$) is normalized and encoded into the amplitudes of a quantum pure state using two $n$-qubit registers R and C, where R indexes the rows and C indexes the columns of the density matrix A. Meanwhile, an auxiliary register $\phi$ is initialized to store the quantum state of optimization weight coefficients. This step completes the mapping from classical data to the quantum system.
	
	Step 2: \textbf{Quantum Superposition and Construction of Correlated States}.
	The initial state is transformed by an operator $W^{(0)}$ composed of Hadamard gates, resulting in a uniform superposition state. Subsequently, the operator $W^{(1)}$ applies a series of controlled operations, introducing index states and correlating them with weight coefficients, thereby initializing the terms in the objective function summation within the quantum state.
	
	Step 3: \textbf{Application of Variational Unitary Transformations}.
	The core step involves applying the operator $W^{(2)}$, which introduces unitary transformations $U(\alpha)$ and $U(\beta)$ controlled by parameters $\alpha$ and $\beta$. These transformations act on the quantum circuit in a controlled manner, effectively equivalent to computing $U^T(\alpha) A U(\beta)$ at the classical level. These parameterized unitary matrices serve as candidate solutions for the singular vectors to be determined.
	
	Step 4: \textbf{Computation of the Objective Function in the Quantum State}.
	The operator $W^{(3)}$ performs controlled matrix multiplication operations, followed by the operator $W^{(4)}$ implementing the final Hadamard transformation. The quantum circuit successfully encodes the value of the objective function into the state amplitude of an auxiliary qubit ($K$). At this stage, the computation of the objective function is completed in one shot within the quantum superposition state.
	
	Step 5: \textbf{Efficient Quantum Measurement and Information Extraction}.
	To read out the objective function value, an efficient measurement scheme is designed: after applying specific basis transformations to the informative qubit ($K$) and an auxiliary qubit ($B$), a joint measurement is performed. By statistically analyzing four simple measurement outcome probabilities, the value of the objective function $L(\alpha, \beta)$ can be precisely computed through classical post-processing.
	
	The quantum component is responsible for efficiently computing the objective function and its gradients, while the classical component iteratively updates the parameters $(\alpha, \beta)$ to maximize $L(\alpha, \beta)$. Upon convergence, the optimal parameters $U(\alpha^*)$ and $U(\beta^*)$ approximate the singular vectors of matrix $A$, and the singular values can be derived from the final objective function value.

	\subsection {Quantum Circuit Realization of the PPT and CCNR Criteria}
	\subsubsection {Quantum Circuit Realization of the PPT Criterion}
		This paper constructs a PPT criterion circuit for an $n$-qubit bipartite state $\rho_{AB}$, where system A is an $m$-qubit system and system B is an $n-m$-qubit system.
		
		To perform the partial transpose operation on system B, the matrix must first be partitioned into blocks. We have developed an initialization method that enables the initial state to already be in the form of a block matrix. First, the density matrix of the unknown quantum state is re-encoded, with the elements of the density matrix encoded into the superposition states of $C_1$, $C_2$, $R_1$, and $R_2$.
		
		The algorithm consists of the following three steps:
	
	\noindent \textbf{Step 1: Constructing the Initial State of the Entire System}
		
		First, introduce two m-qubit registers $C_1$ and $R_1$, and two $(n-m)$-qubit registers $C_2$ and $R_2$, where $R_1$ and $R_2$ index the rows of the density matrix, and $C_1$ and $C_2$ index the columns of the density matrix.
		
		The elements of the quantum state $\rho_{AB}$ are encoded to obtain the following pure state:
	
	\begin{equation}
	 |\pi_{0}\rangle = \sum_{r_1=0}^{2^m-1} \sum_{r_2=0}^{2^{n-m}-1} \sum_{c_1=0}^{2^m-1} \sum_{c_2=0}^{2^{n-m}-1} a_{(r_1 \cdot 2^{n-m} + r_2, c_1 \cdot2^{n-m}+ c_2)} |r_1\rangle_{R_1} |r_2\rangle_{R_2} |c_1\rangle_{C_1}|c_2\rangle_{C_2}
	\end{equation}
	
	\begin{equation}
		\sum_{r_1=0}^{2^m-1} \sum_{r_2=0}^{2^{n-m}-1} \sum_{c_1=0}^{2^m-1} \sum_{c_2=0}^{2^{n-m}-1} a^2_{(r_1 \cdot 2^{n-m} + r_2, c_1 \cdot2^{n-m}+ c_2)}=1
	\end{equation}
	(Specifically, an explanation is provided for the subscript of $a_{(r_1 \cdot 2^{n-m} + r_2, c_1 \cdot 2^{n-m} + c_2)}$: the original matrix is $2^n \times 2^n$, but is now treated as a block matrix with $2^m$ blocks, where each block has dimensions $2^{n-m} \times 2^{n-m}$.)
	
	The circuit for the partial transpose operation is as follows:
	
	\begin{center}
		\begin{tikzpicture}[very thick]
			\draw (0,0) -- (6,0);
			\draw (0,-1) -- (6,-1); 
			\draw (0,-2) -- (6,-2);
			\draw (0,-3) -- (6,-3);
			
			\draw (0.5,0.2) -- (0.4,-0.2);
			\node[above right] at (0.5,0) {$m$};
			\node[left] at (0,0) {$R_1$};
			
			\draw (0.5,-0.8) -- (0.4,-1.2);
			\node[above right] at (0.1,-1) {$n-m$};
			\node[left] at (0,-1) {$R_2$};
			
			\draw (0.5,-1.8) -- (0.4,-2.2);
			\node[above right] at (0.5,-2) {$m$};
			\node[left] at (0,-2) {$C_1$};
			
			\draw (0.5,-2.8) -- (0.4,-3.2);
			\node[above right] at (0.1,-3) {$n-m$};
			\node[left] at (0,-3) {$C_2$};
			
			\draw (3,-1) -- (3,-3);
			\node at (3,-1) {$\boldsymbol{\times}$};
			\node at (3,-3) {$\boldsymbol{\times}$};

			\node[right] at (6.5,-0.5) {$R_T$};
			\draw (6.5,0.2) -- (6.5,-1.2);
			\node[right] at (6.5,-2.5) {$C_T$};
			\draw (6.5,-1.8) -- (6.5,-3.2);
			
			\node[above] at (2,0.2) {$|\pi_0\rangle$};
			\node[above] at (4,0.2) {$|\pi_1\rangle$};
			\draw[red, dashed] (2,0.2) -- (2,-3.4);
			\draw[red, dashed] (4,0.2) -- (4,-3.4);
			
		\end{tikzpicture}
	\end{center}
	where $R_T$ is the total row index after transposition, and $C_T$ is the total column index after transposition.

	\noindent \textbf{Step 2: The states between the two registers $R_2$ and $C_2$ are swapped using SWAP gates.}
	
	\begin{equation}
	|\pi_{1}\rangle =\mathrm{SWAP}_{R_2,C_2} |\pi_{0}\rangle =\sum_{r_1=0}^{2^m-1} \sum_{r_2=0}^{2^{n-m}-1} \sum_{c_1=0}^{2^m-1} \sum_{c_2=0}^{2^{n-m}-1} a_{(r_1 \cdot 2^{n-m} + r_2, c_1 \cdot2^{n-m}+ c_2)} |r_1\rangle_{R_1} |c_2\rangle_{R_2} |c_1\rangle_{C_1}|r_2\rangle_{C_2}
	\end{equation}
	
	\noindent At this point, $|\pi_{1}\rangle$ is the pure state representation of $\rho_{AB}^{T_B}$ in equation (2).
	
	\noindent \textbf{Step 3: Interface with the Singular Value Decomposition Circuit}
	
	The state $|\pi_1\rangle$ is used as input and fed into the trace norm estimation procedure described in Section 3.1.
	The objective function is defined as:
	\begin{equation}
		L(\alpha, \beta)=\sum_j 1/\sqrt{N} \times \text{Re}(\langle j|U(\alpha)^T \rho_{AB}^{T_B} U(\beta)|j\rangle)
	\end{equation}
	After iterations by the classical optimizer and through numerical simulations, we obtain the converged objective function value $L(\alpha, \beta)$. The objective function $L(\alpha, \beta)$ is measured to converge to its maximum value $L(\alpha^*, \beta^*)$. However, unlike in reference \cite{Zenchuk2025}, the input matrix in this work is a density matrix, which already satisfies the normalization condition $\|A\|_F = 1$. Thus, the final computed result of the trace norm is:
	\begin{equation}
		\|\rho_{AB}^{T_B}\| = L(\alpha^*, \beta^*) \times \sqrt{N}
	\end{equation}
	Finally, if $\|\rho_{AB}^{T_B}\| > 1$, the quantum state must be entangled.

	\subsubsection{Quantum Circuit Implementation of CCNR Criterion}
		
		The section constructs the quantum circuit for the CCNR matrix realignment criterion, targeting an $n$-qubit bipartite state $\rho_{AB}$. Since the indices of systems A and B need to be interchanged, the dimensions of systems A and B must be consistent. System A is an $n/2$-qubit system, and system B is also an $n/2$-qubit system.
		
		To perform the matrix realignment operation, the matrix must first be partitioned into blocks. We use the same initialization method as in the previous section, first re-encoding the density matrix of the unknown quantum state by encoding the elements of the density matrix into the superposition states of $C_1$, $C_2$, $R_1$, and $R_2$.
		
		\noindent \textbf{Step 1: Constructing the Initial State of the Entire System}
		
		First, introduce four $N/2$-qubit registers $C_1$, $R_1$, $C_2$, and $R_2$, where $C_1$ and $R_1$ index the rows of the density matrix, and $C_2$ and $R_2$ index the columns of the density matrix.
		
		\noindent The elements of the quantum state $\rho_{AB}$ are encoded to obtain the following pure state:
	
	\begin{equation}
	 |\pi_{0}\rangle = \sum_{r_1=0}^{2^{n/2}-1} \sum_{r_2=0}^{2^{n/2}-1} \sum_{c_1=0}^{2^{n/2}-1} \sum_{c_2=0}^{2^{n/2}-1} a_{(r_1 \cdot 2^{n/2} + r_2, c_1 \cdot2^{n/2}+ c_2)} |r_1\rangle_{R_1} |r_2\rangle_{R_2} |c_1\rangle_{C_1}|c_2\rangle_{C_2}
	\end{equation}
	\begin{equation}
	\sum_{r_1=0}^{2^{n/2}-1} \sum_{r_2=0}^{2^{n/2}-1} \sum_{c_1=0}^{2^{n/2}-1} \sum_{c_2=0}^{2^{n/2}-1} a_{(r_1 \cdot 2^{n/2} + r_2, c_1 \cdot2^{n/2}+ c_2)}
	\end{equation}	
	
	\noindent The quantum circuit for the matrix realignment criterion is as follows:
	
	\begin{center}
		\begin{tikzpicture}[very thick]
			\draw (0,0) -- (6,0);
			\draw (0,-1) -- (6,-1); 
			\draw (0,-2) -- (6,-2);
			\draw (0,-3) -- (6,-3);
			
			\draw (0.5,0.2) -- (0.4,-0.2);
			\node[above right] at (0.2,0) {$n/2$};
			\node[left] at (0,0) {$R_1$};
			
			\draw (0.5,-0.8) -- (0.4,-1.2);
			\node[above right] at (0.2,-1) {$n/2$};
			\node[left] at (0,-1) {$R_2$};
			
			\draw (0.5,-1.8) -- (0.4,-2.2);
			\node[above right] at (0.2,-2) {$n/2$};
			\node[left] at (0,-2) {$C_1$};
			
			\draw (0.5,-2.8) -- (0.4,-3.2);
			\node[above right] at (0.2,-3) {$n/2$};
			\node[left] at (0,-3) {$C_2$};
			
			\draw (3,-1) -- (3,-2);
			\node at (3,-1) {$\boldsymbol{\times}$};
			\node at (3,-2) {$\boldsymbol{\times}$};

			\node[right] at (6.5,-0.5) {$R_R$};
			\draw (6.5,0.2) -- (6.5,-1.2);
			\node[right] at (6.5,-2.5) {$C_R$};
			\draw (6.5,-1.8) -- (6.5,-3.2);
			
			\node[above] at (2,0.2) {$|\pi_0\rangle$};
			\node[above] at (4,0.2) {$|\pi_1\rangle$};
			\draw[red, dashed] (2,0.2) -- (2,-3.4);
			\draw[red, dashed] (4,0.2) -- (4,-3.4);
			
		\end{tikzpicture}
	\end{center}
	where $R_T$ is the total row index after realignment, and $C_T$ is the total column index after realignment.

	\noindent \textbf{Step 2: The states between the two registers $R_2$ and $C_1$ are swapped using SWAP gates.}
	
	\begin{equation}
	|\pi_{1}\rangle = \mathrm{SWAP}_{R_2,C_1} |\pi_{0}\rangle = \sum_{r_1=0}^{2^{n/2}-1} \sum_{r_2=0}^{2^{n/2}-1} \sum_{c_1=0}^{2^{n/2}-1} \sum_{c_2=0}^{2^{n/2}-1} a_{(r_1 \cdot 2^{n/2} + r_2, c_1 \cdot2^{n/2}+ c_2)} |r_1\rangle_{R_1} |c_1\rangle_{R_2} |r_2\rangle_{C_1}|c_2\rangle_{C_2}
	\end{equation}
	\noindent At this point, $|\pi_{1}\rangle$ is the pure state representation of $R(\rho_{AB})$ in equation (2).
	
	\noindent \textbf{Step 3: Interface with the Singular Value Decomposition Circuit}
	
    The state $|\pi_1\rangle$ is used as input and fed into the trace norm estimation procedure described in Section 3.1.  
    The objective function is defined as:
    \begin{equation}
    	L(\alpha, \beta)=\sum_j 1/\sqrt{N} \times \text{Re}(\langle j|U(\alpha)^T R(\rho_{AB}) U(\beta)|j\rangle)
    \end{equation}
    After iterations by the classical optimizer and through numerical simulations, we obtain the converged objective function value $L(\alpha, \beta)$. The objective function $L(\alpha, \beta)$ is measured to converge to its maximum value $L(\alpha^*, \beta^*)$. Thus, the final computed result of the trace norm is:
    \begin{equation}
    	\|R(\rho_{AB})\| = L(\alpha^*, \beta^*) \times \sqrt{N}
    \end{equation}
    Finally, if $\|R(\rho_{AB})\| > 1$, the quantum state must be entangled.

	\subsection{Concrete Practical Example}
	
	Consider the two-qubit Bell state $|\Phi^+\rangle = \frac{1}{\sqrt{2}}(|00\rangle + |11\rangle)$, whose density matrix is:
	
	\begin{equation}
		\rho_{AB} = |\Phi^+\rangle\langle\Phi^+| = \frac{1}{2}(|00\rangle\langle00| + |00\rangle\langle11| + |11\rangle\langle00| + |11\rangle\langle11|)=\frac{1}{2}
		\begin{bmatrix}
			1 & 0 & 0 & 1 \\
			0 & 0 & 0 & 0 \\
			0 & 0 & 0 & 0 \\
			1 & 0 & 0 & 1
		\end{bmatrix}.
	\end{equation}
	
	\noindent Its partial transpose matrix $\rho_{AB}^{T_B}$ and realigned matrix $R(\rho_{AB})$ can be respectively written as:
	
	\begin{equation}
		\rho_{AB}^{T_B} =  \frac{1}{2}(|00\rangle\langle00| + |10\rangle\langle01| + |01\rangle\langle10| + |11\rangle\langle11|) =\frac{1}{2}
		\begin{bmatrix}
			1 & 0 & 0 & 0 \\
			0 & 0 & 1 & 0 \\
			0 & 1 & 0 & 0 \\
			0 & 0 & 0 & 1
		\end{bmatrix}.
	\end{equation}
	
	\noindent and:
	
	\begin{equation}
		R(\rho_{AB})=  \frac{1}{2}(|00\rangle\langle00| + |10\rangle\langle10| + |01\rangle\langle01| + |11\rangle\langle11|) =\frac{1}{2}
		\begin{bmatrix}
			1 & 0 & 0 & 0 \\
			0 & 1 & 0 & 0 \\
			0 & 0 & 1 & 0 \\
			0 & 0 & 0 & 1
		\end{bmatrix}.
	\end{equation}

	\subsubsection{Detection via the PPT Criterion}
		
		\noindent \textbf{Step 1: Constructing the Initial State of the Entire System}
		
		\noindent Encode $\rho_{AB}$ in (20) as follows:
	\begin{equation}
		|\pi_{0}\rangle =\sum_{r_1=0}^{1} \sum_{r_2=0}^{1} \sum_{c_1=0}^{1} \sum_{c_2=0}^{1}  a_{(r_1 \cdot 2+r_2, c_1 \cdot2^+ c_2)} |r_1\rangle_{R_1} |r_2\rangle_{R_2} |c_1\rangle_{C_1}|c_2\rangle_{C_2}
	\end{equation}
	\begin{equation}
		\sum_{r_1=0}^{1} \sum_{r_2=0}^{1} \sum_{c_1=0}^{1} \sum_{c_2=0}^{1} a^2_{(r_1 \cdot 2+r_2, c_1 \cdot2^+ c_2)} =1
	\end{equation}
	
	\noindent The quantum circuit for the PPT criterion at this stage is as follows:
	
	\begin{center}
		\begin{tikzpicture}[very thick]
			\draw (0,0) -- (6,0);
			\draw (0,-1) -- (6,-1); 
			\draw (0,-2) -- (6,-2);
			\draw (0,-3) -- (6,-3);
			
			
			\node[left] at (0,0) {$R_1$};

			\node[left] at (0,-1) {$R_2$};

			\node[left] at (0,-2) {$C_1$};

			\node[left] at (0,-3) {$C_2$};
			
			\draw (3,-1) -- (3,-3);
			\node at (3,-1) {$\boldsymbol{\times}$};
			\node at (3,-3) {$\boldsymbol{\times}$};

			\node[right] at (6.5,-0.5) {$R_T$};
			\draw (6.5,0.2) -- (6.5,-1.2);
			\node[right] at (6.5,-2.5) {$C_T$};
			\draw (6.5,-1.8) -- (6.5,-3.2);
			
			\node[above] at (2,0.2) {$|\pi_0\rangle$};
			\node[above] at (4,0.2) {$|\pi_1\rangle$};
			\draw[red, dashed] (2,0.2) -- (2,-3.4);
			\draw[red, dashed] (4,0.2) -- (4,-3.4);
			
		\end{tikzpicture}
	\end{center}
	where $R_T$ is the total row index after transposition, and $C_T$ is the total column index after transposition.
	
	\noindent \textbf{Step 2: Swap the states between registers $R_2$ and $C_2$ via SWAP gates}
	
	\begin{equation}
		|\pi_{1}\rangle =\mathrm{SWAP}_{R_2,C_2} |\pi_{0}\rangle = \sum_{r_1=0}^{1} \sum_{r_2=0}^{1} \sum_{c_1=0}^{1} \sum_{c_2=0}^{1} a_{(r_1 \cdot 2+r_2, c_1 \cdot 2 + c_2)} |r_1\rangle_{R_1}|c_2\rangle_{R_2} |c_1\rangle_{C_1}|r_2\rangle_{C_2}
	\end{equation}
	
	\noindent \textbf{Step 3: Interface with the Singular Value Decomposition Circuit}
	
	The state $|\pi_1\rangle$ is used as input and fed into the trace norm estimation procedure described in Section 3.1.
	The objective function is defined as:
	\begin{equation}
		L(\alpha, \beta)=\sum_j 1/2 \times \text{Re}(\langle j|U(\alpha)^T \rho_{AB}^{T_B} U(\beta)|j\rangle)
	\end{equation}
	After iterations by the classical optimizer and through numerical simulations, we obtain the converged objective function value $L(\alpha, \beta)$. The objective function $L(\alpha, \beta)$ is measured to converge to its maximum value $L(\alpha^*, \beta^*)=1.0$. Thus, the final computed result of the trace norm is:
	\begin{equation}
		\|\rho_{AB}^{T_B}\| = L(\alpha^*, \beta^*) \times 2
	\end{equation}
	Finally, the value of $\|\rho_{AB}^{T_B}\|$ is 2.
	
	Meanwhile, for intuitive understanding and verification, we perform singular value decomposition on $\rho_{AB}^{T_B}$ as $\rho_{AB}^{T_B} = U D V^\dagger$. Its form is:
	\begin{equation}
		\rho_{AB}^{T_B} = U D V^\top,
	\end{equation}
	where
	\begin{equation}
		U = \begin{bmatrix}
			1 & 0 & 0 & 0 \\
			0 & \frac{1}{\sqrt{2}} & 0 & \frac{1}{\sqrt{2}} \\
			0 & \frac{1}{\sqrt{2}} & 0 & -\frac{1}{\sqrt{2}} \\
			0 & 0 & 1 & 0
		\end{bmatrix}, \quad
		D = \frac{1}{2} I_4, \quad
		V = \begin{bmatrix}
			1 & 0 & 0 & 0 \\
			0 & \frac{1}{\sqrt{2}} & 0 & -\frac{1}{\sqrt{2}} \\
			0 & \frac{1}{\sqrt{2}} & 0 & \frac{1}{\sqrt{2}} \\
			0 & 0 & 1 & 0
		\end{bmatrix}.
	\end{equation}
	
	\noindent All four singular values are $\frac{1}{2}$, and the trace norm is:
	\begin{equation}
		\|\rho_{AB}^{T_B}\| = 2.
	\end{equation}
	The final nuclear norm calculation result is $2$, which is consistent with the output value. We successfully determine that the Bell state $|\Phi^+\rangle$ is entangled. This example fully validates the correctness and feasibility of the entire workflow of our scheme, from quantum state encoding and operations to the final entanglement detection.
	
	\subsubsection{Detection via the CCNR Criterion}
	
	\noindent \textbf{Step 1: Constructing the Initial State of the Entire System}
	
	\noindent Encode $\rho_{AB}$ in (20) as follows:
	\begin{equation}
		|\pi_{0}\rangle = \sum_{r_1=0}^{1} \sum_{r_2=0}^{1} \sum_{c_1=0}^{1} \sum_{c_2=0}^{1} a_{(r_1 \cdot 2+r_2, c_1 \cdot2^+ c_2)} |r_1\rangle_{R_1} |r_2\rangle_{R_2} |c_1\rangle_{C_1}|c_2\rangle_{C_2}
	\end{equation}
	\begin{equation}
	\sum_{r_1=0}^{1} \sum_{r_2=0}^{1} \sum_{c_1=0}^{1} \sum_{c_2=0}^{1} a^2_{(r_1 \cdot 2+r_2, c_1 \cdot2^+ c_2)} =1
	\end{equation}
	
	\noindent The quantum circuit for the CCNR criterion at this stage is as follows:
	
	\begin{center}
		\begin{tikzpicture}[very thick]
			\draw (0,0) -- (6,0);
			\draw (0,-1) -- (6,-1); 
			\draw (0,-2) -- (6,-2);
			\draw (0,-3) -- (6,-3);
			
			\node[left] at (0,0) {$R_1$};

			\node[left] at (0,-1) {$R_2$};

			\node[left] at (0,-2) {$C_1$};

			\node[left] at (0,-3) {$C_2$};
			
			\draw (3,-1) -- (3,-2);
			\node at (3,-1) {$\boldsymbol{\times}$};
			\node at (3,-2) {$\boldsymbol{\times}$};

			\node[right] at (6.5,-0.5) {$R_R$};
			\draw (6.5,0.2) -- (6.5,-1.2);
			\node[right] at (6.5,-2.5) {$C_R$};
			\draw (6.5,-1.8) -- (6.5,-3.2);
			
			\node[above] at (2,0.2) {$|\pi_0\rangle$};
			\node[above] at (4,0.2) {$|\pi_1\rangle$};
			\draw[red, dashed] (2,0.2) -- (2,-3.4);
			\draw[red, dashed] (4,0.2) -- (4,-3.4);
			
		\end{tikzpicture}
	\end{center}
	where $R_T$ is the total row index after realignment, and $C_T$ is the total column index after realignment.
	
	\noindent \textbf{Step 2: Swap the states between registers $R_2$ and $C_1$ via SWAP gates}

	\begin{equation}
		|\pi_{1}\rangle =\mathrm{SWAP}_{R_2,C_1} |\pi_{0}\rangle = \sum_{r_1=0}^{1} \sum_{r_2=0}^{1} \sum_{c_1=0}^{1} \sum_{c_2=0}^{1} a_{(r_1 \cdot 2+r_2, c_1 \cdot2^+ c_2)} |r_1\rangle_{R_1} |c_1\rangle_{R_2} |r_2\rangle_{C_1}|c_2\rangle_{C_2}
	\end{equation}

	\noindent \textbf{Step 3: Interface with the Singular Value Decomposition Circuit}
		
		The state $|\pi_1\rangle$ is used as input and fed into the trace norm estimation procedure described in Section 3.1.
		The objective function is defined as:
		\begin{equation}
			L(\alpha, \beta)=\sum_j 1/2 \times \text{Re}(\langle j|U(\alpha)^T \rho_{AB} U(\beta)|j\rangle)
		\end{equation}
		After iterations by the classical optimizer and through numerical simulations, we obtain the converged objective function value $L(\alpha, \beta)$. The objective function $L(\alpha, \beta)$ is measured to converge to its maximum value $L(\alpha^, \beta^)=1.0$. Thus, the final computed result of the trace norm is:
		\begin{equation}
			\|R(\rho{AB})\| = L(\alpha^*, \beta^*) \times 2
		\end{equation}
		Finally, the value of $\|R(\rho_{AB})\|$ is 2.
		
		Meanwhile, for intuitive understanding and verification, we perform singular value decomposition on $R(\rho_{AB})$ as $R(\rho_{AB}) = U D V^\dagger$. Its form is:
	
	\begin{equation}
	R(\rho_{AB})= U D V^\top,
	\end{equation}
	where
	\begin{equation}
		U = \begin{bmatrix}
			1 & 0 & 0 & 0 \\
			0 & 1 & 0 & 0 \\
			0 & 0 & 1 & 0 \\
			0 & 0 & 0 & 1
		\end{bmatrix}, \quad
		D = \frac{1}{2} I_4, \quad
		V = \begin{bmatrix}
			1 & 0 & 0 & 0 \\
			0 & 1 & 0 & 0 \\
			0 & 0 & 1 & 0 \\
			0 & 0 & 0 & 1
		\end{bmatrix}.
	\end{equation}
	
	\noindent The trace norm is:
	\begin{equation}
		\| R(\rho_{AB}) \| = 2.
	\end{equation}
	The final trace norm calculation result is $2$, which is consistent with the output value. We successfully determine that the Bell state $|\Phi^+\rangle$ is entangled. This example fully validates the correctness and feasibility of the entire workflow of our scheme, from quantum state encoding and operations to the final entanglement detection.

   \section{Conclusion}
	This study systematically designs and theoretically verifies a comprehensive quantum circuit scheme for implementing the PPT and CCNR entanglement criteria, establishing a clear and feasible technical pathway for efficient and direct detection of quantum entanglement. The core contribution of this scheme lies in three key innovations: First, we construct a systematic quantum computational workflow that ingeniously transforms the complex linear algebraic operations of partial transpose and matrix realignment into equivalent, executable quantum operations through the introduction of controlled SWAP gates applied between specific registers (R and C), thereby effectively simulating the partial transpose and realignment of the density matrix at the quantum level. Second, we successfully integrate the resulting quantum state $|\pi_1\rangle$obtained after partial transpose or realignment as input into a specifically adapted VQSVD subroutine. By setting all weighting parameters $q_j$ in the subroutine to be equal, we repurpose its functionality from resolving individual singular values to directly estimating the trace norm of the partially transposed matrix $\rho_{AB}^{T_B}$ or the realigned matrix $R(\rho_{AB})$. Through a quantum-classical hybrid optimization process, the scheme outputs an estimate of the trace norm, thereby providing a decisive criterion for the presence of entanglement.
	
	The implications of this work are far-reaching. Theoretically, it transforms a highly abstract mathematical criterion into a concrete and operable quantum algorithm blueprint, significantly advancing the qualitative analysis of quantum entanglement toward quantitative and computable directions. Technically, the scheme demonstrates how to deeply integrate matrix encoding, controlled logic gates, and variational hybrid algorithms: a paradigm that can be widely extended to other quantum information processing tasks requiring analysis of matrix spectral properties, such as quantum channel discrimination and state fidelity estimation, thus offering important methodological value. Finally, in terms of application prospects, with the continuous development of intermediate-scale noisy quantum computing technology, this scheme provides a powerful theoretical tool and implementation pathway for the direct and efficient analysis and certification of entanglement structures in complex quantum systems on future quantum devices. Its potential advantages will become particularly prominent when handling large-scale quantum states beyond the reach of classical computational capabilities, thereby opening new possibilities for the application of quantum computing in fundamental physics research and advanced information science.

	\renewcommand{\refname}{References}

\end{document}